\documentclass[12pt,a4paper,english,aps,superscriptaddress,groupedaddress,showpacs,showkeys]{revtex4}

\usepackage[T1]{fontenc}
\usepackage[latin9]{inputenc}
\usepackage{babel}
\usepackage{amsmath}
\usepackage{amssymb}
\usepackage{esint}
\usepackage[unicode=true]
 {hyperref}
\usepackage{breakurl}

\makeatletter

\@ifundefined{textcolor}{}
{%
 \definecolor{BLACK}{gray}{0}
 \definecolor{WHITE}{gray}{1}
 \definecolor{RED}{rgb}{1,0,0}
 \definecolor{GREEN}{rgb}{0,1,0}
 \definecolor{BLUE}{rgb}{0,0,1}
 \definecolor{CYAN}{cmyk}{1,0,0,0}
 \definecolor{MAGENTA}{cmyk}{0,1,0,0}
 \definecolor{YELLOW}{cmyk}{0,0,1,0}
}

\usepackage{dcolumn}
\usepackage{bm}
\usepackage{amsfonts}

\makeatother

\begin{document}

\title{General pseudo self-adjoint boundary conditions for a 1D KFG particle
in a box}

\author{Salvatore De Vincenzo}

\homepage{https://orcid.org/0000-0002-5009-053X}

\email{[salvatored@nu.ac.th]}

\affiliation{The Institute for Fundamental Study (IF), Naresuan University, Phitsanulok
65000, Thailand}

\date{February 23, 2023}
\begin{abstract}
\noindent \textbf{Abstract} We consider a 1D Klein-Fock-Gordon particle
in a finite interval, or box. We construct for the first time the
most general set of pseudo self-adjoint boundary conditions for the
Hamiltonian operator that is present in the first order in time 1D
Klein-Fock-Gordon wave equation, or the 1D Feshbach-Villars wave equation.
We show that this set depends on four real parameters and can be written
in terms of the one-component wavefunction for the second order in
time 1D Klein-Fock-Gordon wave equation and its spatial derivative,
both evaluated at the endpoints of the box. Certainly, we write the
general set of pseudo self-adjoint boundary conditions also in terms
of the two-component wavefunction for the 1D Feshbach-Villars wave
equation and its spatial derivative, evaluated at the ends of the
box; however, the set actually depends on these two column vectors
each multiplied by the singular matrix that is present in the kinetic
energy term of the Hamiltonian. As a consequence, we found that the
two-component wavefunction for the 1D Feshbach-Villars equation and
its spatial derivative do not necessarily satisfy the same boundary
condition that these quantities satisfy when multiplied by the singular
matrix. In any case, given a particular boundary condition for the
one-component wavefunction of the standard 1D Klein-Fock-Gordon equation
and using the pair of relations that arise from the very definition
of the two-component wavefunction for the 1D Feshbach-Villars equation,
the respective boundary condition for the latter wavefunction and
its derivative can be obtained. Our results can be extended to the
problem of a 1D Klein-Fock-Gordon particle moving on a real line with
a point interaction (or a hole) at one point.
\end{abstract}



\maketitle

\section{Introduction}

\noindent As is well known, the three-dimensional (3D) Klein-Fock-Gordon
(KFG) wave equation in its standard form plays an important role in
relativistic quantum mechanics \cite{RefA,RefB,RefC,RefD,RefE}. As
an example, when potentials fail to create particle-antiparticle pairs,
the 3D KFG wave equation can be used to describe spin-zero particles,
for example, the pion, a composite particle, and the Higgs boson,
an apparently elementary particle. Clearly, this equation is one of
the most widely used in relativistic quantum mechanics. Naturally,
the search for exact solutions to this equation in specific and representative
potentials has always been of interest, mainly because these solutions
can be useful for modeling real physical processes. In the study of
exactly solvable problems, various methods have been introduced and
developed. Examples include supersymmetric quantum mechanics (SUSY
QM) and/or the factorization method \cite{RefF,RefG,RefH,RefI,RefJ}
and the Nikiforov-Uvarov (UV) method \cite{RefH,RefI,RefK}, among
others \cite{RefL,RefM,RefN,RefO}. It is worth mentioning that in
recent years, new computational schemes or methods have been applied
to obtain solutions of nonlinear partial differential equations that
are related in some way to the KFG equation. See, for example, Refs.
\cite{RefP,RefQ,RefR} and references therein. 

In reviewing the literature on KFG theory, it is immediately apparent
that the 3D KFG wave equation in Hamiltonian form, i.e., the so-called
3D Feshbach-Villars (FV) wave equation \cite{RefS}, has not received
the same attention as the standard 3D KFG equation. Certainly, both
equations are equivalent, and connecting their corresponding solutions
seems to be straightforward. However, the 3D FV partial differential
equation is first order in time and second order in space, that is,
it includes a second-order Hamiltonian operator in the spatial derivative
(for a nice discussion of the procedure used by Feshbach and Villars
to obtain a linear equation in the time derivative, see Ref. \cite{RefT}.
For a brief and concise historical discussion of similar work, but
prior to that of Feshbach and Villars, see again Ref. \cite{RefT},
specifically, the commentary written in its reference number 3, page
191). 

Similarly, the one-dimensional (1D) FV wave equation has also not
received sufficient attention when considering problems within the
KFG theory in (1+1) dimensions. Certainly, the 1D KFG equation in
its standard form is much more popular. In this regard, there is an
issue within the 1D KFG theory that has received practically no attention
and that we can raise with the following questions: What are the boundary
conditions that the 1D FV equation can support? Can general families
of boundary conditions be written for this equation? Specifically,
what are the appropriate boundary conditions for this equation in
the problem of a 1D KFG particle inside an interval? For example,
some unexpected boundary conditions for the solutions of the 1D FV
wave equation in simple physical situations were presented in Refs.
\cite{RefU,RefV,RefW}. In general, the boundary conditions for the
solutions of the second-order KFG equation in 3D and 1D appear to
be similar to those supported by the corresponding Schr\"{o}dinger wavefunction
(see, for example, Refs. \cite{RefU,RefW,RefX,RefY}), but we do not
have at our disposal a wave equation that could have boundary conditions
similar to those of the 1D FV equation (the presence of a singular
matrix in the kinetic energy term of the Hamiltonian has much to do
with this). In general, the physically acceptable boundary conditions
for a wave equation that is written in Hamiltonian form must ensure
that the respective Hamiltonian operator retains its essential attribute,
namely, that of being self-adjoint (if that is the case). In the case
of the 1D FV equation, it is known that its Hamiltonian is a formally
pseudo-Hermitian operator (or a formally pseudo self-adjoint operator)
\cite{RefB,RefD}, and, in principle, we could find families of general
boundary conditions that agree with the property of being a pseudo
self-adjoint operator, i.e., not just formally. In fact, here, we
show that indeed a general four-parameter family of boundary conditions
can be found for the solutions of the 1D FV equation and that it is
consistent with the latter property. Incidentally, to do this is essentially
to specify the domain of the Hamiltonian and that of its generalized
adjoint (as is done in the case of Hamiltonians that are self-adjoint
in the standard way), but, in addition, these two domains must be
equal, i.e., they must always contain the same boundary condition
(once the four parameters are fixed). 

The article is organized as follows. In Section II, we begin by introducing
the KFG equations in their standard and Hamiltonian versions and the
relations linking their solutions. In addition, we introduce the pseudo
inner product for the two-component solutions of the 1D FV equation
and briefly discuss its relation to other distinctive inner products
of quantum mechanics. In particular, we note that this pseudo inner
product can also be considered the scalar product for the one-component
solutions of the KFG equation in its standard form. Moreover, as might
be expected, this pseudo inner product does not possess the property
of positive definiteness but can be independent of time. Thus, the
corresponding pseudo norm can be a constant, and because this implies
that the probability current density takes the same value at each
end of the box, the Hamiltonian for this problem can be a pseudo-Hermitian
operator. In fact, the Hamiltonian is formally pseudo-Hermitian, and
we find in this section a general four-parameter set of boundary conditions
that ensures that it is indeed a pseudo-Hermitian operator. We write
this set in terms of the one-component wavefunction for the 1D KFG
wave equation and its spatial derivative, both evaluated at the ends
of the interval. Here, we also consider the nonrelativistic approximation
of the general set of boundary conditions, and the results support
the idea that this set is indeed the most general. In Section III,
we finally write the general set of boundary conditions in terms of
the two-component column vector for the 1D FV wave equation and its
spatial derivative, evaluated at the ends of the interval. To be precise,
the set must be written in terms of the latter two column vectors
each multiplied by the singular matrix that is present in the kinetic
energy term of the Hamiltonian (remember that a singular matrix does
not have an inverse). In Section IV (Appendix I), we check that the
time derivative of the pseudo inner product of two solutions of the
1D FV equation in a nonzero electric potential, but expressed in terms
of the respective solutions of the standard KFG equation in the same
potential, is proportional to a term evaluated at the ends of the
box that also does not depend on the potential, i.e., it is a boundary
term. In Section V (Appendix II), we show that the Hamiltonian operator
for a 1D KFG particle in a box is in fact a pseudo self-adjoint operator;
that is, the general matrix boundary condition, i.e., the general
set of boundary conditions, ensures that the domains of the Hamiltonian
and its generalized adjoint are equal. From the results shown in this
section, it follows that the boundary term that arose in Section IV
(Appendix I) always vanishes (certainly, for any boundary condition
included in the general family of boundary conditions); consequently,
the value of the pseudo inner product in this problem is conserved.
Finally, concluding remarks are presented in Section VI. 

\section{Boundary conditions for the 1D KFG particle in a box I}

Let us begin by writing the 1D KFG wave equation in Hamiltonian form,
\begin{equation}
\mathrm{i}\hbar\frac{\partial}{\partial t}\Psi=\hat{\mathrm{h}}\Psi,
\end{equation}
where 
\begin{equation}
\hat{\mathrm{h}}=-\frac{\hbar^{2}}{2\mathrm{m}}\left(\hat{\tau}_{3}+\mathrm{i}\hat{\tau}_{2}\right)\frac{\partial^{2}}{\partial x^{2}}+\mathrm{m}c^{2}\hat{\tau}_{3}+V(x)\hat{1}_{2},
\end{equation}
is, let us say, the KFG Hamiltonian differential operator. Here, $\hat{\tau}_{3}=\hat{\sigma}_{z}$
and $\hat{\tau}_{2}=\hat{\sigma}_{y}$ are Pauli matrices and $V(x)\in\mathbb{R}$
is the external electric potential ($\hat{1}_{2}$ is the $2\times2$
identity matrix). The (matrix) operator $\hat{\mathrm{h}}$ acts on
(complex) two-component column state vectors of the form $\Psi=\Psi(x,t)=\left[\,\psi_{1}(x,t)\;\,\psi_{2}(x,t)\,\right]^{\mathrm{T}}$
(the symbol $^{\mathrm{T}}$ represents the transpose of a matrix).
Equation (1) with $\hat{\mathrm{h}}$ given in Eq. (2) is the 1D FV
wave equation \cite{RefB,RefC,RefD,RefS}.

The 1D KFG wave equation in its standard form, or the second order
in time KFG equation in one spatial dimension \cite{RefA,RefE} is
given by 
\begin{equation}
\left[\,\mathrm{i}\hbar\frac{\partial}{\partial t}-V(x)\right]^{2}\psi=\left[-\hbar^{2}c^{2}\frac{\partial^{2}}{\partial x^{2}}+(\mathrm{m}c^{2})^{2}\right]\psi,
\end{equation}
where $\psi=\psi(x,t)$ is a (complex) one-component state vector
or one-component wavefunction. 

The relation between $\psi$ and $\Psi$ can be defined as follows:
\begin{equation}
\Psi=\left[\begin{array}{c}
\psi_{1}\\
\psi_{2}
\end{array}\right]=\frac{1}{2}\left[\begin{array}{c}
\psi+\mathrm{i}\tau\left(\frac{\partial}{\partial t}-\frac{V}{\mathrm{i}\hbar}\right)\psi\\
\psi-\mathrm{i}\tau\left(\frac{\partial}{\partial t}-\frac{V}{\mathrm{i}\hbar}\right)\psi
\end{array}\right],
\end{equation}
where $\tau\equiv\hbar/\mathrm{m}c^{2}$. The Compton wavelength is
precisely $\lambda_{\mathrm{C}}\equiv c\tau$; thus, $\tau$ is the
time taken for a ray of light to travel the distance $\lambda_{\mathrm{C}}$.
The expression given in Eq. (3) is fully equivalent to Eq. (1) (with
$\hat{\mathrm{h}}$ given in Eq. (2)) \cite{RefB,RefC}. Note that,
from Eq. (4), the solution $\psi$ of Eq. (3) depends only on the
components of the column vector $\Psi$, namely, 
\begin{equation}
\psi=\psi_{1}+\psi_{2}.
\end{equation}
Additionally,
\begin{equation}
\left(\mathrm{i}\hbar\frac{\partial}{\partial t}\psi-V\psi\right)\frac{1}{\mathrm{m}c^{2}}=\psi_{1}-\psi_{2}.
\end{equation}

\noindent Certainly, all the results we have presented so far are
well known. 

Let us now consider a 1D KFG particle moving in the interval $x\in\Omega=[a,b]$,
i.e., in a box. The corresponding Hamiltonian operator given in Eq.
(2) acts on two-component column state vectors of the form $\Psi=\left[\,\psi_{1}\;\,\psi_{2}\,\right]^{\mathrm{T}}$
and $\Phi=\left[\,\phi_{1}\;\,\phi_{2}\,\right]^{\mathrm{T}}$, and
the scalar product for these two state vectors must be defined as
\begin{equation}
\langle\langle\Psi,\Phi\rangle\rangle\equiv\int_{\Omega}\mathrm{d}x\,\Psi^{\dagger}\hat{\tau}_{3}\Phi
\end{equation}
(the symbol $^{\dagger}$ denotes the usual Hermitian conjugate, or
the usual formal adjoint, of a matrix and an operator) \cite{RefB,RefC,RefD,RefS}.
Additionally, the square of the corresponding norm (or rather, pseudo
norm) is $\left\Vert \left\Vert \Psi\right\Vert \right\Vert ^{2}\equiv\langle\langle\Psi,\Psi\rangle\rangle=\int_{\Omega}\mathrm{d}x\,\varrho$,
where $\varrho=\varrho(x,t)=\Psi^{\dagger}\hat{\tau}_{3}\Psi=|\psi_{1}|^{2}-|\psi_{2}|^{2}$
is the 1D KFG probability density. Certainly, $\varrho$ is not positive
definite and calling it probability density is not absolutely correct
(although it can be interpreted as a charge density) \cite{RefB,RefC,RefD,RefS}.
Note that the integral in (7) can also be identified with the usual
scalar product in Dirac's theory in (1+1) dimensions, namely, $\langle\Psi,\Phi\rangle_{\mathrm{D}}\equiv\int_{\Omega}\mathrm{d}x\,\Psi^{\dagger}\Phi$,
which is an inner product on the Hilbert space of two-component square-integrable
wavefunctions, $\mathcal{L}^{2}(\Omega)\oplus\mathcal{L}^{2}(\Omega)$;
therefore, 
\begin{equation}
\langle\langle\Psi,\Phi\rangle\rangle\equiv\langle\Psi,\hat{\tau}_{3}\Phi\rangle_{\mathrm{D}},
\end{equation}
and $\langle\Psi,\Phi\rangle_{\mathrm{D}}=\langle\langle\Psi,\hat{\tau}_{3}\Phi\rangle\rangle$.
Because $\langle\langle\Psi,\Psi\rangle\rangle$ can be a negative
quantity, the scalar product in Eq. (7) is an indefinite (or improper)
inner product, or a pseudo inner product, on an infinite-dimensional
complex vector space. In general, such a vector space itself is not
necessarily a Hilbert space.

Similarly, writing $\Psi$ and $\Phi$ in the integrand in (7) in
terms of their respective components, that is, using the relations
that arise from Eq. (4) and other analogous relations for $\Phi$
(which are obtained from Eq. (4) by making the replacements $\Psi\rightarrow\Phi$,
$\psi_{1}\rightarrow\phi_{1}$, $\psi_{2}\rightarrow\phi_{2}$ and
$\psi\rightarrow\phi$), we obtain
\begin{equation}
\langle\langle\Psi,\Phi\rangle\rangle=\frac{\mathrm{i}\hbar}{2\mathrm{m}c^{2}}\int_{\Omega}\mathrm{d}x\,\left(\psi^{*}\phi_{t}-\psi_{t}^{*}\phi-\frac{2V}{\mathrm{i}\hbar}\psi^{*}\phi\right)
\end{equation}

\noindent (where the asterisk $^{*}$ denotes the complex conjugate,
and $\psi_{t}\equiv\partial\psi/\partial t$, etc), or also,
\begin{equation}
\langle\langle\Psi,\Phi\rangle\rangle=\frac{\mathrm{i}\hbar}{2\mathrm{m}c^{2}}\left(\langle\psi,\phi_{t}\rangle_{\mathrm{S}}-\langle\psi_{t},\phi\rangle_{\mathrm{S}}-\frac{2}{\mathrm{i}\hbar}\langle\psi,V\phi\rangle_{\mathrm{S}}\right)\equiv\langle\psi,\phi\rangle_{\mathrm{KFG}},
\end{equation}
where $\langle\psi,\phi\rangle_{\mathrm{KFG}}$ can be considered
the scalar product for the one-component solutions of the 1D KFG equation
in Eq. (3) (see Appendix I). Note that $\langle\;,\;\rangle_{\mathrm{S}}$
denotes the usual scalar product in the Schr\"{o}dinger theory in one
spatial dimension, namely, $\langle\psi,\phi\rangle_{\mathrm{S}}\equiv\int_{\Omega}\mathrm{d}x\,\psi^{*}\phi$,
which is an inner product on the Hilbert space of one-component square-integrable
wavefunctions, $\mathcal{L}^{2}(\Omega)$. Certainly, $\psi$ and
$\psi_{t}$, and $\phi$, $V\phi$, and $\phi_{t}$, must belong to
$\mathcal{L}^{2}(\Omega)$ to ensure that $\langle\psi,\phi\rangle_{\mathrm{KFG}}$
exists \cite{RefZ}. 

It can be noted that there is an isomorphism between the vectorial
space of the solutions $\psi$ of the standard 1D KFG equation for
the corresponding 1D particle, namely, 
\begin{equation}
\left[\left(\partial_{t}-\frac{V}{\mathrm{i}\hbar}\right)^{2}+\hat{\mathrm{d}}\,\right]\psi=0
\end{equation}
(Eq. (3)), where $\hat{\mathrm{d}}\equiv-c^{2}\partial_{xx}+\tau^{-2}$
($\partial_{t}\equiv\partial/\partial t$ and $\partial_{xx}\equiv\partial^{2}/\partial x^{2}$,
etc) and the vectorial space of the initial state vectors of the 1D
KFG equation in Hamiltonian form for this 1D particle, namely, Eq.
(1) with $\hat{\mathrm{h}}$ given in Eq. (2) \cite{RefAA}. In effect,
a possible initial state vector, for example, at $t=0$, would have
the form 
\begin{equation}
\Psi(0)=\left[\begin{array}{c}
\psi_{1}(0)\\
\psi_{2}(0)
\end{array}\right]=\frac{1}{2}\left[\begin{array}{c}
\psi(0)+\mathrm{i}\tau\left(\psi_{t}(0)-\frac{V}{\mathrm{i}\hbar}\psi(0)\right)\\
\psi(0)-\mathrm{i}\tau\left(\psi_{t}(0)-\frac{V}{\mathrm{i}\hbar}\psi(0)\right)
\end{array}\right],
\end{equation}
that arises immediately from the relation given in Eq. (4). Thus,
giving an initial state vector as $\Psi(0)$ is equivalent to providing
the initial data for the solution vector $\psi$, namely, $\psi(0)$
and $\psi_{t}(0)$. Incidentally, operators $\hat{\mathrm{d}}$, which
can act on the one-component state vectors $\psi$, and \textrm{$\hat{\mathrm{h}}$},
which can act on the two-component state vectors $\Psi$, are related
as follows: 
\begin{equation}
\hat{\mathrm{h}}=+\frac{\hbar}{2}\tau\left(\hat{\tau}_{3}+\mathrm{i}\hat{\tau}_{2}\right)\hat{\mathrm{d}}+\frac{\hbar}{2}\tau^{-1}\left(\hat{\tau}_{3}-\mathrm{i}\hat{\tau}_{2}\right)+V(x)\hat{1}_{2}.
\end{equation}

Although the scalar product in Eqs. (7) and (10) does not possess
the property of positive definiteness (i.e., $\langle\langle\Psi,\Psi\rangle\rangle<0$),
it is a time-independent scalar product. Indeed, using Eq. (3) for
$\psi$ and $\psi^{*}$, and for $\phi$ and $\phi^{*}$, it can be
demonstrated that the following relation is verified: 
\begin{equation}
\frac{\mathrm{d}}{\mathrm{d}t}\langle\langle\Psi,\Phi\rangle\rangle=-\frac{\mathrm{i}\hbar}{2\mathrm{m}}\left.\left[\,\psi_{x}^{*}\,\phi-\psi^{*}\phi_{x}\,\right]\right|_{a}^{b}=\frac{\mathrm{d}}{\mathrm{d}t}\langle\psi,\phi\rangle_{\mathrm{KFG}},
\end{equation}
where $\left.\left[\, g\,\right]\right|_{a}^{b}\equiv g(b,t)-g(a,t)$,
and $\psi_{x}\equiv\partial\psi/\partial x$, etc. This result is
also valid when the external potential $V$ is different from zero
inside the box (see Appendix I). The term evaluated at the endpoints
of the interval $\Omega$ must vanish due to the boundary condition
satisfied by $\psi$ and $\phi$, or $\Psi$ and $\Phi$ (see Appendix
II). Additionally, if we make $\psi=\phi$, or $\Psi=\Phi$, in Eq.
(14), we obtain the result 
\begin{equation}
\frac{\mathrm{d}}{\mathrm{d}t}\langle\langle\Psi,\Psi\rangle\rangle=-\left.\left[\, j\,\right]\right|_{a}^{b}=\frac{\mathrm{d}}{\mathrm{d}t}\langle\psi,\psi\rangle_{\mathrm{KFG}},
\end{equation}
where $j=j(x,t)=(\mathrm{i}\hbar/2\mathrm{m})(\psi_{x}^{*}\,\psi-\psi^{*}\psi_{x})$
would be the probability current density, although we know that this
quantity, as well as $\varrho$, cannot be interpreted as probability
quantities \cite{RefB,RefC}. The disappearance of the boundary term
in Eq. (15) implies that the pseudo norm remains constant, and because
$j(a,t)=j(b,t)$, we have that $\hat{\mathrm{h}}$ must be a pseudo-Hermitian
operator. In the case that $\Omega=\mathbb{R}$, the scalar product
$\langle\langle\Psi,\Phi\rangle\rangle$ is a time-independent constant
whenever $\Psi$ and $\Phi$ are two normalizable solutions, i.e.,
solutions that have their pseudo norm finite. The square of the pseudo
norm of these functions could be negative, but their magnitude cannot
be infinite if the boundary term in Eq. (14) is expected to be zero.

Next, we use the pseudo inner product given in Eq. (7), which is defined
over an indefinite inner product space \cite{RefT}. For a collection
of basic properties of this scalar product (but also of general results
on Hamiltonians of the type given in Eq. (2)), see Ref. \cite{RefAA}.
Using integration by parts twice, it can be demonstrated that the
Hamiltonian differential operator $\hat{\mathrm{h}}$ in Eq. (2) satisfies
the following relation: 
\begin{equation}
\langle\langle\Psi,\hat{\mathrm{h}}\Phi\rangle\rangle=\langle\langle\hat{\mathrm{h}}_{\mathrm{adj}}\Psi,\Phi\rangle\rangle+f[\Psi,\Phi],
\end{equation}
where the boundary term $f[\Psi,\Phi]$ is given by 
\begin{equation}
f[\Psi,\Phi]\equiv\frac{\hbar^{2}}{2\mathrm{m}}\left.\left[\,\Psi_{x}^{\dagger}\,\hat{\tau}_{3}\,(\hat{\tau}_{3}+\mathrm{i}\hat{\tau}_{2})\Phi-\Psi^{\dagger}\,\hat{\tau}_{3}\,(\hat{\tau}_{3}+\mathrm{i}\hat{\tau}_{2})\Phi_{x}\,\right]\right|_{a}^{b}.
\end{equation}
This quantity can also be written in a way that will be especially
important, namely,
\begin{equation}
f[\Psi,\Phi]\equiv\frac{\hbar^{2}}{2\mathrm{m}}\,\frac{1}{2}\left.\left[\,\left((\hat{\tau}_{3}+\mathrm{i}\hat{\tau}_{2})\Psi_{x}\right)^{\dagger}(\hat{\tau}_{3}+\mathrm{i}\hat{\tau}_{2})\Phi-\left((\hat{\tau}_{3}+\mathrm{i}\hat{\tau}_{2})\Psi\right)^{\dagger}(\hat{\tau}_{3}+\mathrm{i}\hat{\tau}_{2})\Phi_{x}\,\right]\right|_{a}^{b}.
\end{equation}
The latter somewhat unexpected expression is true because the singular
matrix $\hat{\tau}_{3}+\mathrm{i}\hat{\tau}_{2}$ obeys the following
relation: $(\hat{\tau}_{3}+\mathrm{i}\hat{\tau}_{2})^{\dagger}(\hat{\tau}_{3}+\mathrm{i}\hat{\tau}_{2})=2\hat{\tau}_{3}\,(\hat{\tau}_{3}+\mathrm{i}\hat{\tau}_{2})$;
however, $(\hat{\tau}_{3}+\mathrm{i}\hat{\tau}_{2})^{2}=\hat{0}$.
The differential operator $\hat{\mathrm{h}}_{\mathrm{adj}}$ in Eq.
(16) is the generalized Hermitian conjugate, or the formal generalized
adjoint of $\hat{\mathrm{h}}$, namely,
\begin{equation}
\hat{\mathrm{h}}_{\mathrm{adj}}=\hat{\eta}^{-1}\,\hat{\mathrm{h}}^{\dagger}\,\hat{\eta}=\hat{\tau}_{3}\,\hat{\mathrm{h}}^{\dagger}\,\hat{\tau}_{3}
\end{equation}
($\hat{\eta}=\hat{\tau}_{3}=\hat{\eta}^{-1}$ is sometimes called
the metric operator; in this case, $\hat{\eta}$ is a bounded operator
and satisfies $\hat{\eta}^{3}=\hat{\eta}$) and therefore (just formally,
i.e., by using only the scalar product definition given in Eq. (7)),
\begin{equation}
\langle\langle\Psi,\hat{\mathrm{h}}\Phi\rangle\rangle=\langle\langle\hat{\mathrm{h}}_{\mathrm{adj}}\Psi,\Phi\rangle\rangle.
\end{equation}
The latter is essentially the relation that defines the generalized
adjoint differential operator $\hat{\mathrm{h}}_{\mathrm{adj}}$ on
an indefinite inner product space. Clearly, the latter definition
requires that $f[\Psi,\Phi]$ in Eq. (16) vanishes. 

The Hamiltonian operator in Eq. (2) also formally satisfies the following
relation: 
\begin{equation}
\hat{\mathrm{h}}=\hat{\mathrm{h}}_{\mathrm{adj}},
\end{equation}
that is, $\hat{\mathrm{h}}$ is formally pseudo-Hermitian (or formally
generalized Hermitian), or formally pseudo self-adjoint (or formally
generalized self-adjoint). However, if the boundary conditions imposed
on $\Psi$ and $\Phi$ at the endpoints of the interval $\Omega$
lead to the cancellation of the boundary term in Eq. (16), then the
differential operator $\hat{\mathrm{h}}$ is indeed pseudo-Hermitian
(or generalized Hermitian), and as shown in Appendix II, it is also
pseudo self-adjoint (or generalized self-adjoint), i.e., 
\begin{equation}
\langle\langle\Psi,\hat{\mathrm{h}}\Phi\rangle\rangle=\langle\langle\hat{\mathrm{h}}\Psi,\Phi\rangle\rangle.
\end{equation}
Precisely, we want to obtain a general set of boundary conditions
for the pseudo-Hermitian Hamiltonian differential operator. Thus,
if we impose $\Psi=\Phi$ in the latter relation and in Eq. (16) (with
the result in Eq. (21)), we obtain the following condition:
\begin{equation}
f[\Psi,\Psi]=\frac{\hbar}{\mathrm{i}}\left.\left[\, j\,\right]\right|_{a}^{b}=0\quad\left(\;\Rightarrow\; j(b,t)=j(a,t)\,\right),
\end{equation}
where $j=j(x,t)$ is given by 
\begin{equation}
j=\frac{\mathrm{i}\hbar}{2\mathrm{m}}\,\frac{1}{2}\left[\,\left((\hat{\tau}_{3}+\mathrm{i}\hat{\tau}_{2})\Psi_{x}\right)^{\dagger}(\hat{\tau}_{3}+\mathrm{i}\hat{\tau}_{2})\Psi-\left((\hat{\tau}_{3}+\mathrm{i}\hat{\tau}_{2})\Psi\right)^{\dagger}(\hat{\tau}_{3}+\mathrm{i}\hat{\tau}_{2})\Psi_{x}\,\right]
\end{equation}
(see Eq. (18)). But also because $\hat{\tau}_{3}\,(\hat{\tau}_{3}+\mathrm{i}\hat{\tau}_{2})=\hat{1}_{2}+\hat{\sigma}_{x}$
(the latter if we use the expression given by Eq. (17)), and the result
in Eq. (5), we obtain
\begin{equation}
j=\frac{\mathrm{i}\hbar}{2\mathrm{m}}\left(\,\psi_{x}^{*}\,\psi-\psi^{*}\psi_{x}\,\right),
\end{equation}
as expected (see the comment made just after Eq. (15)). Certainly,
all the generalized Hermitian boundary conditions must lead to the
equality of $j$ at the endpoints of the interval $\Omega$. Furthermore,
we also obtain the result $\langle\langle\Psi,\hat{\mathrm{h}}\Psi\rangle\rangle=\langle\langle\hat{\mathrm{h}}\Psi,\Psi\rangle\rangle=\langle\langle\Psi,\hat{\mathrm{h}}\Psi\rangle\rangle^{*}$
(the superscript $^{*}$ denotes the complex conjugate); therefore,
$\langle\langle\Psi,\hat{\mathrm{h}}\Psi\rangle\rangle\equiv\langle\langle\hat{\mathrm{h}}\rangle\rangle_{\Psi}\in\mathbb{R}$,
i.e., the generalized mean value of the Hamiltonian operator is real
valued. Other typical properties of operators that are Hermitian in
the usual sense hold here as well; for example, the eigenvalues are
real (see, for example, Refs. \cite{RefB,RefD}). 

Substituting $j$ from Eq. (25) into Eq. (23), we obtain the result
(we omit the variable $t$ in the expressions that follow) 
\[
\lambda\frac{2\mathrm{m}}{\hbar^{2}}f[\Psi,\Psi]=\left.\left[\,\psi\,\lambda\psi_{x}^{*}-\psi^{*}\lambda\psi_{x}\,\right]\right|_{a}^{b}
\]
\begin{equation}
=\left[\,\psi(b)\,\lambda\psi_{x}^{*}(b)-\psi^{*}(b)\,\lambda\psi_{x}(b)\,\right]-\left[\,\psi(a)\,\lambda\psi_{x}^{*}(a)-\psi^{*}(a)\,\lambda\psi_{x}(a)\,\right]=0,
\end{equation}
where $\lambda\in\mathbb{R}$ is a parameter required for dimensional
reasons. It is very convenient to rewrite the latter two terms using
the following identity:
\[
z_{1}z_{2}^{*}-z_{1}^{*}z_{2}=\frac{\mathrm{i}}{2}\left[\,(z_{1}+\mathrm{i}z_{2})(z_{1}+\mathrm{i}z_{2})^{*}-(z_{1}-\mathrm{i}z_{2})(z_{1}-\mathrm{i}z_{2})^{*}\,\right]
\]
\begin{equation}
=\frac{\mathrm{i}}{2}\left(\,\left|z_{1}+\mathrm{i}z_{2}\right|^{2}-\left|z_{1}-\mathrm{i}z_{2}\right|^{2}\,\right),
\end{equation}
 where $z_{1}$ and $z_{2}$ are complex numbers. Then, the following
result is obtained: 
\[
\lambda\frac{2\mathrm{m}}{\hbar^{2}}f[\Psi,\Psi]=\frac{\mathrm{i}}{2}\left(\,\left|\psi(b)+\mathrm{i}\lambda\psi_{x}(b)\right|^{2}-\left|\psi(b)-\mathrm{i}\lambda\psi_{x}(b)\right|^{2}\,\right)
\]
\[
-\frac{\mathrm{i}}{2}\left(\,\left|\psi(a)+\mathrm{i}\lambda\psi_{x}(a)\right|^{2}-\left|\psi(a)-\mathrm{i}\lambda\psi_{x}(a)\right|^{2}\,\right)
\]
\[
=\frac{\mathrm{i}}{2}\left(\,\left|\psi(b)+\mathrm{i}\lambda\psi_{x}(b)\right|^{2}+\left|\psi(a)-\mathrm{i}\lambda\psi_{x}(a)\right|^{2}\,\right)
\]
\begin{equation}
-\frac{\mathrm{i}}{2}\left(\,\left|\psi(b)-\mathrm{i}\lambda\psi_{x}(b)\right|^{2}+\left|\psi(a)+\mathrm{i}\lambda\psi_{x}(a)\right|^{2}\,\right)=0,
\end{equation}
that is,
\[
\lambda\frac{2\mathrm{m}}{\hbar^{2}}f[\Psi,\Psi]=\frac{\mathrm{i}}{2}\left[\begin{array}{c}
\psi(b)+\mathrm{i}\lambda\psi_{x}(b)\\
\psi(a)-\mathrm{i}\lambda\psi_{x}(a)
\end{array}\right]^{\dagger}\left[\begin{array}{c}
\psi(b)+\mathrm{i}\lambda\psi_{x}(b)\\
\psi(a)-\mathrm{i}\lambda\psi_{x}(a)
\end{array}\right]
\]
\begin{equation}
-\frac{\mathrm{i}}{2}\left[\begin{array}{c}
\psi(b)-\mathrm{i}\lambda\psi_{x}(b)\\
\psi(a)+\mathrm{i}\lambda\psi_{x}(a)
\end{array}\right]^{\dagger}\left[\begin{array}{c}
\psi(b)-\mathrm{i}\lambda\psi_{x}(b)\\
\psi(a)+\mathrm{i}\lambda\psi_{x}(a)
\end{array}\right]=0.
\end{equation}

Let us now consider the following general matrix boundary condition:
\begin{equation}
\left[\begin{array}{c}
\psi(b)+\mathrm{i}\lambda\psi_{x}(b)\\
\psi(a)-\mathrm{i}\lambda\psi_{x}(a)
\end{array}\right]=\hat{\mathrm{M}}\left[\begin{array}{c}
\psi(b)-\mathrm{i}\lambda\psi_{x}(b)\\
\psi(a)+\mathrm{i}\lambda\psi_{x}(a)
\end{array}\right],
\end{equation}
where $\hat{\mathrm{M}}$ is an arbitrary complex matrix. By substituting
Eq. (30) into Eq. (29), we obtain
\[
\frac{\mathrm{i}}{2}\left[\begin{array}{c}
\psi(b)-\mathrm{i}\lambda\psi_{x}(b)\\
\psi(a)+\mathrm{i}\lambda\psi_{x}(a)
\end{array}\right]^{\dagger}\left(\hat{\mathrm{M}}^{\dagger}\hat{\mathrm{M}}-\hat{1}_{2}\right)\left[\begin{array}{c}
\psi(b)-\mathrm{i}\lambda\psi_{x}(b)\\
\psi(a)+\mathrm{i}\lambda\psi_{x}(a)
\end{array}\right]=0;
\]
therefore, $\hat{\mathrm{M}}$ is a unitary matrix (the justification
for this result is given in the comment that follows Eq. (A14)). Thus,
a general set of generalized Hermitian boundary conditions for the
1D KFG particle in a box can be written as follows:
\begin{equation}
\left[\begin{array}{c}
\psi(b)-\mathrm{i}\lambda\psi_{x}(b)\\
\psi(a)+\mathrm{i}\lambda\psi_{x}(a)
\end{array}\right]=\hat{\mathrm{U}}_{(2\times2)}\left[\begin{array}{c}
\psi(b)+\mathrm{i}\lambda\psi_{x}(b)\\
\psi(a)-\mathrm{i}\lambda\psi_{x}(a)
\end{array}\right],
\end{equation}
where $\hat{\mathrm{U}}_{(2\times2)}=\hat{\mathrm{M}}^{-1}$ is also
unitary. This family of boundary conditions is similar to the one
corresponding to the problem of the 1D Schr\"{o}dinger particle enclosed
in a box; for example, see Eq. (28) in Ref. \cite{RefBB}. In relation
to this, we can also take the nonrelativistic approximation of the
general boundary condition given in Eq. (31). For that purpose, it
is convenient to first write the KFG wavefunction $\psi=\psi(x,t)$
as follows: $\psi=\psi_{\mathrm{S}}\exp(-\mathrm{i}\,\mathrm{m}c^{2}t/\hbar)$,
where $\psi_{\mathrm{S}}=\psi_{\mathrm{S}}(x,t)$ is the Schr\"{o}dinger
wavefunction. Because in this approximation we have that $\left|\,\mathrm{i}\hbar(\psi_{\mathrm{S}})_{t}\,\right|\ll\mathrm{m}c^{2}\left|\,\psi_{\mathrm{S}}\,\right|$,
we can write $\psi_{t}=(-\mathrm{i}\,\mathrm{m}c^{2}t/\hbar)\psi$,
and therefore $\psi_{1}=\left(1-\frac{V}{2\mathrm{m}c^{2}}\right)\psi$
and $\psi_{2}=\frac{V}{2\mathrm{m}c^{2}}\psi$ (see Eq. (4)). Thus,
for weak external potentials and to the lowest order in $v/c$ (and
for positive energy solutions), $\psi_{1}\approx\psi$ satisfies the
Schr\"{o}dinger equation in the potential $V+\mathrm{m}c^{2}$ (the latter
$\mathrm{m}c^{2}$ can be eliminated by using the expression $\psi_{1}\approx\psi=\psi_{\mathrm{S}}\exp(-\mathrm{i}\,\mathrm{m}c^{2}t/\hbar)$)
but also $(\psi_{1})_{x}\approx\psi_{x}$ (see, for example, Refs.
\cite{RefB,RefS,RefW}). It is then clear that, in the problem of
the particle in a box, the one-component KFG wavefunction satisfies
the same boundary conditions as the one-component Schr\"{o}dinger wavefunction.
Incidentally, a similar result to Eq. (31) had already been obtained
by taking the nonrelativistic limit of the most general family of
boundary conditions for the 1D Dirac particle enclosed in a box \cite{RefCC}.
Additionally, in the analogous problem of a 1D Schr\"{o}dinger particle
in the presence of a point interaction at the point $x=0$ (or a hole
at the origin), the most general family of boundary conditions is
similar to that given in Eq. (31) \cite{RefDD}. Indeed, all these
results substantiate that the set of boundary conditions dependent
on the four real parameters given in Eq. (31) is also the most general
for a 1D KFG particle in the interval $[a,b]$. Moreover, by making
the replacements $a\rightarrow0+$ and $b\rightarrow0-$ in Eq. (31),
we obtain the respective most general set of boundary conditions for
the case in which the 1D KFG particle moves along the real line with
a hole at the origin. Some examples of boundary conditions for this
system can be seen in Refs. \cite{RefU,RefW} and will be briefly
discussed in Section III. 

For all the boundary conditions that are part of the general set of
boundary conditions in Eq. (31), $\hat{\mathrm{h}}$ is a pseudo-Hermitian
operator, but it is also a pseudo self-adjoint operator (see Appendix
II). Certainly, the result in Eq. (31) is given in terms of the wavefunction
$\psi$, but if the relation in Eq. (5) is used, it can also be written
in terms of the components of $\Psi=\left[\,\psi_{1}\;\,\psi_{2}\,\right]^{\mathrm{T}}$,
i.e., in terms of $\psi_{1}+\psi_{2}$, and its spatial derivative
$(\psi_{1})_{x}+(\psi_{2})_{x}$, evaluated at the edges $x=a$ and
$x=b$. Actually, the general family of boundary conditions given
in Eq. (31) must be written in terms of $(\hat{\tau}_{3}+\mathrm{i}\hat{\tau}_{2})\Psi$
and $(\hat{\tau}_{3}+\mathrm{i}\hat{\tau}_{2})\Psi_{x}$ evaluated
at the ends of the box. We work on this in the next section. We give
below some examples of boundary conditions that are contained in Eq.
(31): $\psi(a)=\psi(b)=0$ ($\hat{\mathrm{U}}_{(2\times2)}=-\hat{1}_{2}$),
i.e., $\psi$ can satisfy the Dirichlet boundary condition; $\psi_{x}(a)=\psi_{x}(b)=0$
($\hat{\mathrm{U}}_{(2\times2)}=+\hat{1}_{2}$), i.e., $\psi$ can
satisfy the Neumann boundary condition; $\psi(a)=\psi(b)$ and $\psi_{x}(a)=\psi_{x}(b)$
($\hat{\mathrm{U}}_{(2\times2)}=+\hat{\sigma}_{x}$), $\psi$ can
satisfy the periodic boundary condition; $\psi(a)=-\psi(b)$ and $\psi_{x}(a)=-\psi_{x}(b)$
($\hat{\mathrm{U}}_{(2\times2)}=-\hat{\sigma}_{x}$), $\psi$ can
satisfy the antiperiodic boundary condition; $\psi(a)=\psi_{x}(b)=0$
($\hat{\mathrm{U}}_{(2\times2)}=\hat{\sigma}_{z}$), i.e., $\psi$
can satisfy a mixed boundary condition; $\psi_{x}(a)=\psi(b)=0$ ($\hat{\mathrm{U}}_{(2\times2)}=-\hat{\sigma}_{z}$),
i.e., $\psi$ can satisfy another mixed boundary condition; $\psi(a)-\lambda\psi_{x}(a)=0$
and $\psi(b)+\lambda\psi_{x}(b)=0$ ($\hat{\mathrm{U}}_{(2\times2)}=\mathrm{i}\hat{1}_{2}$),
$\psi$ can satisfy a kind of Robin boundary condition. In fact, the
latter boundary condition would be the KFG version of the boundary
condition commonly used in the so-called (one-dimensional) MIT bag
model for hadronic structures (see, for example, Ref. \cite{RefCC}).
All these boundary conditions are typical of wave equations that are
of the second order in the spatial derivative.

Of all the boundary conditions included in the four-parameter family
of boundary conditions, only those arising from a diagonal unitary
matrix describe a particle in an impenetrable box. This is because,
for these boundary conditions, the probability current density satisfies
the relation $j(b)=j(a)=0$ for all $t$. Thus, the most general family
of confining boundary conditions for a 1D KFG particle in a box only
has two (real) parameters. The latter result is due to the similarity
between the general set of boundary conditions given in Eq. (31) and
the general sets of boundary conditions for the 1D Dirac and Schr\"{o}dinger
particles, and because we already know that the confining boundary
conditions come from a matrix $\hat{\mathrm{U}}_{(2\times2)}$ that
is diagonal \cite{RefCC}. 

\section{\noindent Boundary conditions for the 1D KFG particle in a box II}

\noindent Here, we obtain the most general set of pseudo self-adjoint
boundary conditions for the Hamiltonian operator in the 1D FV equation,
that is, we write the latter set in terms of $\Psi$ and $\Psi_{x}$
evaluated at the endpoints of the box. More specifically, in terms
of $(\hat{\tau}_{3}+\mathrm{i}\hat{\tau}_{2})\Psi$ and $(\hat{\tau}_{3}+\mathrm{i}\hat{\tau}_{2})\Psi_{x}$.
Indeed, following a procedure similar to that used above to obtain
Eq. (26), namely, substituting $j$ from Eq. (24) into Eq. (23), we
obtain \textrm{
\[
\lambda\frac{2\mathrm{m}}{\hbar^{2}}f[\Psi,\Psi]=\frac{1}{2}\left.\left[\,\left((\hat{\tau}_{3}+\mathrm{i}\hat{\tau}_{2})\lambda\Psi_{x}\right)^{\dagger}(\hat{\tau}_{3}+\mathrm{i}\hat{\tau}_{2})\Psi-\left((\hat{\tau}_{3}+\mathrm{i}\hat{\tau}_{2})\Psi\right)^{\dagger}(\hat{\tau}_{3}+\mathrm{i}\hat{\tau}_{2})\lambda\Psi_{x}\,\right]\right|_{a}^{b}
\]
\[
=\frac{1}{2}\left[\,\left((\hat{\tau}_{3}+\mathrm{i}\hat{\tau}_{2})\lambda\Psi_{x}(b)\right)^{\dagger}(\hat{\tau}_{3}+\mathrm{i}\hat{\tau}_{2})\Psi(b)-\left((\hat{\tau}_{3}+\mathrm{i}\hat{\tau}_{2})\Psi(b)\right)^{\dagger}(\hat{\tau}_{3}+\mathrm{i}\hat{\tau}_{2})\lambda\Psi_{x}(b)\,\right]
\]
\begin{equation}
-\frac{1}{2}\left[\,\left((\hat{\tau}_{3}+\mathrm{i}\hat{\tau}_{2})\lambda\Psi_{x}(a)\right)^{\dagger}(\hat{\tau}_{3}+\mathrm{i}\hat{\tau}_{2})\Psi(a)-\left((\hat{\tau}_{3}+\mathrm{i}\hat{\tau}_{2})\Psi(a)\right)^{\dagger}(\hat{\tau}_{3}+\mathrm{i}\hat{\tau}_{2})\lambda\Psi_{x}(a)\,\right]=0,
\end{equation}
}where again, we insert the real parameter $\lambda$ for dimensional
reasons. Now, we use the following matrix identity twice:
\begin{equation}
\hat{\mathrm{Z}}_{2}^{\dagger}\,\hat{\mathrm{Z}}_{1}-\hat{\mathrm{Z}}_{1}^{\dagger}\,\hat{\mathrm{Z}}_{2}=\frac{\mathrm{i}}{2}\left[\,(\hat{\mathrm{Z}}_{1}+\mathrm{i}\hat{\mathrm{Z}}_{2})^{\dagger}(\hat{\mathrm{Z}}_{1}+\mathrm{i}\hat{\mathrm{Z}}_{2})-(\hat{\mathrm{Z}}_{1}-\mathrm{i}\hat{\mathrm{Z}}_{2})^{\dagger}(\hat{\mathrm{Z}}_{1}-\mathrm{i}\hat{\mathrm{Z}}_{2})\,\right].
\end{equation}
Then, we obtain the following result:
\[
\lambda\frac{2\mathrm{m}}{\hbar^{2}}f[\Psi,\Psi]=\frac{1}{2}\frac{\mathrm{i}}{2}\left[\,\left((\hat{\tau}_{3}+\mathrm{i}\hat{\tau}_{2})(\Psi+\mathrm{i}\lambda\Psi_{x})(b)\right)^{\dagger}(\hat{\tau}_{3}+\mathrm{i}\hat{\tau}_{2})(\Psi+\mathrm{i}\lambda\Psi_{x})(b)\right.
\]
\[
\left.-\left((\hat{\tau}_{3}+\mathrm{i}\hat{\tau}_{2})(\Psi-\mathrm{i}\lambda\Psi_{x})(b)\right)^{\dagger}(\hat{\tau}_{3}+\mathrm{i}\hat{\tau}_{2})(\Psi-\mathrm{i}\lambda\Psi_{x})(b)\,\right]
\]
\[
-\frac{1}{2}\frac{\mathrm{i}}{2}\left[\,\left((\hat{\tau}_{3}+\mathrm{i}\hat{\tau}_{2})(\Psi+\mathrm{i}\lambda\Psi_{x})(a)\right)^{\dagger}(\hat{\tau}_{3}+\mathrm{i}\hat{\tau}_{2})(\Psi+\mathrm{i}\lambda\Psi_{x})(a)\right.
\]
\begin{equation}
\left.-\left((\hat{\tau}_{3}+\mathrm{i}\hat{\tau}_{2})(\Psi-\mathrm{i}\lambda\Psi_{x})(a)\right)^{\dagger}(\hat{\tau}_{3}+\mathrm{i}\hat{\tau}_{2})(\Psi-\mathrm{i}\lambda\Psi_{x})(a)\,\right]=0,
\end{equation}
that is,
\[
\lambda\frac{2\mathrm{m}}{\hbar^{2}}f[\Psi,\Psi]=\frac{1}{2}\frac{\mathrm{i}}{2}\left[\begin{array}{c}
(\hat{\tau}_{3}+\mathrm{i}\hat{\tau}_{2})(\Psi+\mathrm{i}\lambda\Psi_{x})(b)\\
(\hat{\tau}_{3}+\mathrm{i}\hat{\tau}_{2})(\Psi-\mathrm{i}\lambda\Psi_{x})(a)
\end{array}\right]^{\dagger}\left[\begin{array}{c}
(\hat{\tau}_{3}+\mathrm{i}\hat{\tau}_{2})(\Psi+\mathrm{i}\lambda\Psi_{x})(b)\\
(\hat{\tau}_{3}+\mathrm{i}\hat{\tau}_{2})(\Psi-\mathrm{i}\lambda\Psi_{x})(a)
\end{array}\right]
\]
\begin{equation}
-\frac{1}{2}\frac{\mathrm{i}}{2}\left[\begin{array}{c}
(\hat{\tau}_{3}+\mathrm{i}\hat{\tau}_{2})(\Psi-\mathrm{i}\lambda\Psi_{x})(b)\\
(\hat{\tau}_{3}+\mathrm{i}\hat{\tau}_{2})(\Psi+\mathrm{i}\lambda\Psi_{x})(a)
\end{array}\right]^{\dagger}\left[\begin{array}{c}
(\hat{\tau}_{3}+\mathrm{i}\hat{\tau}_{2})(\Psi-\mathrm{i}\lambda\Psi_{x})(b)\\
(\hat{\tau}_{3}+\mathrm{i}\hat{\tau}_{2})(\Psi+\mathrm{i}\lambda\Psi_{x})(a)
\end{array}\right]=0.
\end{equation}

Now, we propose writing a general matrix boundary condition as follows:
\begin{equation}
\left[\begin{array}{c}
(\hat{\tau}_{3}+\mathrm{i}\hat{\tau}_{2})(\Psi+\mathrm{i}\lambda\Psi_{x})(b)\\
(\hat{\tau}_{3}+\mathrm{i}\hat{\tau}_{2})(\Psi-\mathrm{i}\lambda\Psi_{x})(a)
\end{array}\right]=\hat{\mathrm{A}}\left[\begin{array}{c}
(\hat{\tau}_{3}+\mathrm{i}\hat{\tau}_{2})(\Psi-\mathrm{i}\lambda\Psi_{x})(b)\\
(\hat{\tau}_{3}+\mathrm{i}\hat{\tau}_{2})(\Psi+\mathrm{i}\lambda\Psi_{x})(a)
\end{array}\right],
\end{equation}
where $\hat{\mathrm{A}}$ is an arbitrary $4\times4$ complex matrix.
By substituting Eq. (36) into Eq. (35), we obtain
\[
\frac{1}{2}\frac{\mathrm{i}}{2}\left[\begin{array}{c}
(\hat{\tau}_{3}+\mathrm{i}\hat{\tau}_{2})(\Psi-\mathrm{i}\lambda\Psi_{x})(b)\\
(\hat{\tau}_{3}+\mathrm{i}\hat{\tau}_{2})(\Psi+\mathrm{i}\lambda\Psi_{x})(a)
\end{array}\right]^{\dagger}\left(\hat{\mathrm{A}}^{\dagger}\hat{\mathrm{A}}-\hat{1}_{4}\right)\left[\begin{array}{c}
(\hat{\tau}_{3}+\mathrm{i}\hat{\tau}_{2})(\Psi-\mathrm{i}\lambda\Psi_{x})(b)\\
(\hat{\tau}_{3}+\mathrm{i}\hat{\tau}_{2})(\Psi+\mathrm{i}\lambda\Psi_{x})(a)
\end{array}\right]=0,
\]
then $\hat{\mathrm{A}}$ is a unitary matrix ($\hat{1}_{4}$ is the
$4\times4$ identity matrix). Note that the components of the column
vectors in Eq. (36) are themselves $2\times1$ column matrices and
are given by
\begin{equation}
(\hat{\tau}_{3}+\mathrm{i}\hat{\tau}_{2})(\Psi\pm\mathrm{i}\lambda\Psi_{x})(x)=\left[\begin{array}{c}
(\psi\pm\mathrm{i}\lambda\psi_{x})(x)\\
-(\psi\pm\mathrm{i}\lambda\psi_{x})(x)
\end{array}\right]\,,\quad x=a,b.
\end{equation}
Thus, the general boundary condition in Eq. (36) can be written as
follows:
\begin{equation}
\left[\begin{array}{c}
(\psi+\mathrm{i}\lambda\psi_{x})(b)\\
-(\psi+\mathrm{i}\lambda\psi_{x})(b)\\
(\psi-\mathrm{i}\lambda\psi_{x})(a)\\
-(\psi-\mathrm{i}\lambda\psi_{x})(a)
\end{array}\right]=\hat{\mathrm{A}}\left[\begin{array}{c}
(\psi-\mathrm{i}\lambda\psi_{x})(b)\\
-(\psi-\mathrm{i}\lambda\psi_{x})(b)\\
(\psi+\mathrm{i}\lambda\psi_{x})(a)\\
-(\psi+\mathrm{i}\lambda\psi_{x})(a)
\end{array}\right].
\end{equation}
On the other hand, this relation can also be written as follows: 
\begin{equation}
\left[\begin{array}{c}
(\psi+\mathrm{i}\lambda\psi_{x})(b)\\
(\psi-\mathrm{i}\lambda\psi_{x})(a)\\
(\psi+\mathrm{i}\lambda\psi_{x})(b)\\
(\psi-\mathrm{i}\lambda\psi_{x})(a)
\end{array}\right]=\hat{\mathrm{S}}\hat{\mathrm{A}}\hat{\mathrm{S}}^{\dagger}\left[\begin{array}{c}
(\psi-\mathrm{i}\lambda\psi_{x})(b)\\
(\psi+\mathrm{i}\lambda\psi_{x})(a)\\
(\psi-\mathrm{i}\lambda\psi_{x})(b)\\
(\psi+\mathrm{i}\lambda\psi_{x})(a)
\end{array}\right],
\end{equation}
where $\hat{\mathrm{S}}$ is given by
\[
\hat{\mathrm{S}}=\left[\begin{array}{cccc}
1 & 0 & 0 & 0\\
0 & 0 & 1 & 0\\
0 & -1 & 0 & 0\\
0 & 0 & 0 & -1
\end{array}\right]
\]
 
\begin{equation}
=\frac{1}{2}\left(\hat{\sigma}_{z}\otimes\hat{1}_{2}+\mathrm{i}\hat{\sigma}_{y}\otimes\hat{\sigma}_{x}+\mathrm{i}\hat{\sigma}_{x}\otimes\hat{\sigma}_{y}+\hat{1}_{2}\otimes\hat{\sigma}_{z}\right),
\end{equation}
where $\otimes$ denotes the Zehfuss-Kronecker product of matrices,
or the matrix direct product
\begin{equation}
\hat{\mathrm{F}}\otimes\hat{\mathrm{G}}\equiv\left[\begin{array}{ccc}
\mathrm{F}_{11}\hat{\mathrm{G}} & \cdots & \mathrm{F}_{1n}\hat{\mathrm{G}}\\
\vdots & \ddots & \vdots\\
\mathrm{F}_{m1}\hat{\mathrm{G}} & \cdots & \mathrm{F}_{mn}\hat{\mathrm{G}}
\end{array}\right],
\end{equation}
which is bilinear and associative and satisfies, among other properties,
the mixed-product property: $(\hat{\mathrm{F}}\otimes\hat{\mathrm{G}})(\mathrm{\hat{J}}\otimes\hat{\mathrm{K}})=(\hat{\mathrm{F}}\mathrm{\hat{J}}\otimes\hat{\mathrm{G}}\hat{\mathrm{K}})$
(see, for example, Ref. \cite{RefEE}). The matrix $\hat{\mathrm{S}}$
is unitary, and therefore, $\hat{\mathrm{S}}\hat{\mathrm{A}}\hat{\mathrm{S}}^{\dagger}$
is also a unitary matrix. Now, notice that the left-hand side of the
relation in Eq. (39) is given by (see Eq. (30))
\begin{equation}
\left[\begin{array}{c}
\underset{}{\left[\begin{array}{c}
(\psi+\mathrm{i}\lambda\psi_{x})(b)\\
(\psi-\mathrm{i}\lambda\psi_{x})(a)
\end{array}\right]}\\
\left[\begin{array}{c}
(\psi+\mathrm{i}\lambda\psi_{x})(b)\\
(\psi-\mathrm{i}\lambda\psi_{x})(a)
\end{array}\right]
\end{array}\right]=\left[\begin{array}{c}
\underset{}{\hat{\mathrm{M}}\left[\begin{array}{c}
(\psi-\mathrm{i}\lambda\psi_{x})(b)\\
(\psi+\mathrm{i}\lambda\psi_{x})(a)
\end{array}\right]}\\
\hat{\mathrm{M}}\left[\begin{array}{c}
(\psi-\mathrm{i}\lambda\psi_{x})(b)\\
(\psi+\mathrm{i}\lambda\psi_{x})(a)
\end{array}\right]
\end{array}\right]=\left[\begin{array}{cc}
\hat{\mathrm{M}} & \hat{0}\\
\hat{0} & \hat{\mathrm{M}}
\end{array}\right]\left[\begin{array}{c}
(\psi-\mathrm{i}\lambda\psi_{x})(b)\\
(\psi+\mathrm{i}\lambda\psi_{x})(a)\\
(\psi-\mathrm{i}\lambda\psi_{x})(b)\\
(\psi+\mathrm{i}\lambda\psi_{x})(a)
\end{array}\right],
\end{equation}
and substituting the latter relation into Eq. (39), we obtain
\begin{equation}
\hat{\mathrm{S}}\hat{\mathrm{A}}\hat{\mathrm{S}}^{\dagger}=\left[\begin{array}{cc}
\hat{\mathrm{M}} & \hat{0}\\
\hat{0} & \hat{\mathrm{M}}
\end{array}\right]=\hat{1}_{2}\otimes\hat{\mathrm{M}}
\end{equation}
(because $\hat{\mathrm{M}}$ is a unitary matrix, the block diagonal
matrix in Eq. (43) is also unitary). Then, from Eq. (43), we can write
the matrix $\hat{\mathrm{A}}$ as follows: 
\begin{equation}
\hat{\mathrm{A}}=\hat{\mathrm{S}}^{\dagger}\left[\begin{array}{cc}
\hat{\mathrm{M}} & \hat{0}\\
\hat{0} & \hat{\mathrm{M}}
\end{array}\right]\hat{\mathrm{S}}=\hat{\mathrm{S}}^{\dagger}(\hat{1}_{2}\otimes\hat{\mathrm{M}})\,\hat{\mathrm{S}}.
\end{equation}
Thus, the most general family of pseudo self-adjoint boundary conditions
for the 1D KFG particle in a box, that is, for the Hamiltonian operator
in the 1D FV wave equation, can be written as follows (see Eq. (36)):
\begin{equation}
\left[\begin{array}{c}
(\hat{\tau}_{3}+\mathrm{i}\hat{\tau}_{2})(\Psi-\mathrm{i}\lambda\Psi_{x})(b)\\
(\hat{\tau}_{3}+\mathrm{i}\hat{\tau}_{2})(\Psi+\mathrm{i}\lambda\Psi_{x})(a)
\end{array}\right]=\hat{\mathrm{U}}_{(4\times4)}\left[\begin{array}{c}
(\hat{\tau}_{3}+\mathrm{i}\hat{\tau}_{2})(\Psi+\mathrm{i}\lambda\Psi_{x})(b)\\
(\hat{\tau}_{3}+\mathrm{i}\hat{\tau}_{2})(\Psi-\mathrm{i}\lambda\Psi_{x})(a)
\end{array}\right],
\end{equation}
where
\[
\hat{\mathrm{U}}_{(4\times4)}=\hat{\mathrm{A}}^{-1}=\hat{\mathrm{A}}^{\dagger}=\hat{\mathrm{S}}^{\dagger}\left[\begin{array}{cc}
\hat{\mathrm{M}}^{\dagger} & \hat{0}\\
\hat{0} & \hat{\mathrm{M}}^{\dagger}
\end{array}\right]\hat{\mathrm{S}}=\hat{\mathrm{S}}^{\dagger}\left[\begin{array}{cc}
\hat{\mathrm{M}}^{-1} & \hat{0}\\
\hat{0} & \hat{\mathrm{M}}^{-1}
\end{array}\right]\hat{\mathrm{S}}
\]
\begin{equation}
=\hat{\mathrm{S}}^{\dagger}\left[\begin{array}{cc}
\hat{\mathrm{U}}_{(2\times2)} & \hat{0}\\
\hat{0} & \hat{\mathrm{U}}_{(2\times2)}
\end{array}\right]\hat{\mathrm{S}}=\hat{\mathrm{S}}^{\dagger}(\hat{1}_{2}\otimes\hat{\mathrm{U}}_{(2\times2)})\hat{\mathrm{S}}
\end{equation}
(to reach this result, we use Eq. (44) and the fact that $\hat{\mathrm{U}}_{(2\times2)}=\hat{\mathrm{M}}^{-1}$,
the latter two results and only some properties of the matrix direct
product could also be used). Note that the general matrix boundary
condition in Eq. (45) could also be written as follows:
\begin{equation}
(\hat{1}_{2}\otimes(\hat{\tau}_{3}+\mathrm{i}\hat{\tau}_{2}))\left[\begin{array}{c}
(\Psi-\mathrm{i}\lambda\Psi_{x})(b)\\
(\Psi+\mathrm{i}\lambda\Psi_{x})(a)
\end{array}\right]=\hat{\mathrm{U}}_{(4\times4)}(\hat{1}_{2}\otimes(\hat{\tau}_{3}+\mathrm{i}\hat{\tau}_{2}))\left[\begin{array}{c}
(\Psi+\mathrm{i}\lambda\Psi_{x})(b)\\
(\Psi-\mathrm{i}\lambda\Psi_{x})(a)
\end{array}\right];
\end{equation}
however, the matrix $\hat{1}_{2}\otimes(\hat{\tau}_{3}+\mathrm{i}\hat{\tau}_{2})$
does not have an inverse and the column vector on the left side of
this relation cannot be cleared. Thus, the expression given in Eq.
(47) is an elegant way to write the general boundary condition, but
it is not functional and could lead to errors.

The boundary conditions that were presented just before the last paragraph
of Sect. II can be extracted from Eq. (45) if the matrix $\hat{\mathrm{U}}_{(2\times2)}$
is known. In effect, the Dirichlet boundary condition is $(\hat{\tau}_{3}+\mathrm{i}\hat{\tau}_{2})\Psi(a)=(\hat{\tau}_{3}+\mathrm{i}\hat{\tau}_{2})\Psi(b)=0$
($\hat{\mathrm{U}}_{(4\times4)}=-\hat{1}_{4}=-\hat{1}_{2}\otimes\hat{1}_{2}$);
the Neumann boundary condition is $(\hat{\tau}_{3}+\mathrm{i}\hat{\tau}_{2})\Psi_{x}(a)=(\hat{\tau}_{3}+\mathrm{i}\hat{\tau}_{2})\Psi_{x}(b)=0$
($\hat{\mathrm{U}}_{(4\times4)}=+\hat{1}_{4}=+\hat{1}_{2}\otimes\hat{1}_{2}$);
the periodic boundary condition is $(\hat{\tau}_{3}+\mathrm{i}\hat{\tau}_{2})\Psi(a)=(\hat{\tau}_{3}+\mathrm{i}\hat{\tau}_{2})\Psi(b)$
and $(\hat{\tau}_{3}+\mathrm{i}\hat{\tau}_{2})\Psi_{x}(a)=(\hat{\tau}_{3}+\mathrm{i}\hat{\tau}_{2})\Psi_{x}(b)$
($\hat{\mathrm{U}}_{(4\times4)}=\hat{\sigma}_{x}\otimes\hat{1}_{2}$);
the antiperiodic boundary condition is $(\hat{\tau}_{3}+\mathrm{i}\hat{\tau}_{2})\Psi(a)=-(\hat{\tau}_{3}+\mathrm{i}\hat{\tau}_{2})\Psi(b)$
and $(\hat{\tau}_{3}+\mathrm{i}\hat{\tau}_{2})\Psi_{x}(a)=-(\hat{\tau}_{3}+\mathrm{i}\hat{\tau}_{2})\Psi_{x}(b)$
($\hat{\mathrm{U}}_{(4\times4)}=-\hat{\sigma}_{x}\otimes\hat{1}_{2}$);
a mixed boundary condition is $(\hat{\tau}_{3}+\mathrm{i}\hat{\tau}_{2})\Psi(a)=(\hat{\tau}_{3}+\mathrm{i}\hat{\tau}_{2})\Psi_{x}(b)=0$
($\hat{\mathrm{U}}_{(4\times4)}=\hat{\sigma}_{z}\otimes\hat{1}_{2}$);
another mixed boundary condition is $(\hat{\tau}_{3}+\mathrm{i}\hat{\tau}_{2})\Psi_{x}(a)=(\hat{\tau}_{3}+\mathrm{i}\hat{\tau}_{2})\Psi(b)=0$
($\hat{\mathrm{U}}_{(4\times4)}=-\hat{\sigma}_{z}\otimes\hat{1}_{2}$);
a kind of Robin boundary condition (and a kind of MIT bag boundary
condition for a 1D KFG particle) is $(\hat{\tau}_{3}+\mathrm{i}\hat{\tau}_{2})(\Psi(a)-\lambda\Psi_{x}(a))=0$
and $(\hat{\tau}_{3}+\mathrm{i}\hat{\tau}_{2})(\Psi(b)+\lambda\Psi_{x}(b))=0$
($\hat{\mathrm{U}}_{(4\times4)}=\mathrm{i}\hat{1}_{4}=\mathrm{i}\hat{1}_{2}\otimes\hat{1}_{2}$).
Then, to write all these boundary conditions in terms of $\psi(a)$
and $\psi(b)$, and $\psi_{x}(a)$ and $\psi_{x}(b)$, we must use
the fact that $\Psi=\left[\,\psi_{1}\;\,\psi_{2}\,\right]^{\mathrm{T}}$
and $\psi=\psi_{1}+\psi_{2}$ (Eq. (5)). If we wish to obtain explicit
relations between the components of $\Psi$ and $\Psi_{x}$ at $x=a$
and $\Psi$ and $\Psi_{x}$ at $x=b$, we must use the relations given
in Eqs. (5) and (6). Additionally, it can be shown that when the matrix
$\hat{\mathrm{U}}_{(2\times2)}$ is diagonal, then the matrix $\hat{\mathrm{U}}_{(4\times4)}$
is also diagonal; consequently, diagonal matrices $\hat{\mathrm{U}}_{(4\times4)}$
in Eq. (45) lead to confining boundary conditions (see the last paragraph
of Sect. II).

In general, the boundary conditions imposed on $(\hat{\tau}_{3}+\mathrm{i}\hat{\tau}_{2})\Psi$
and $(\hat{\tau}_{3}+\mathrm{i}\hat{\tau}_{2})\Psi_{x}$ at the endpoints
of the box do not imply that $\Psi$ and $\Psi_{x}$ must also satisfy
them. For example, let us consider the problem of the 1D KFG particle
in the step potential ($V(x)=V_{0}\,\Theta(x)$, where $\Theta(x)$
is the Heaviside step function). This problem was also considered
in Refs. \cite{RefU,RefW}. The step potential is a (soft) point interaction
in the neighborhood of the origin, that is, between the points $x=a\rightarrow0+$
and $x=b\rightarrow0-$, and the boundary condition is the periodic
boundary condition, which in this case becomes the continuity condition
of $(\hat{\tau}_{3}+\mathrm{i}\hat{\tau}_{2})\Psi$ and $(\hat{\tau}_{3}+\mathrm{i}\hat{\tau}_{2})\Psi_{x}$
at $x=0$, i.e., $(\hat{\tau}_{3}+\mathrm{i}\hat{\tau}_{2})\Psi(0-)=(\hat{\tau}_{3}+\mathrm{i}\hat{\tau}_{2})\Psi(0+)$
and $(\hat{\tau}_{3}+\mathrm{i}\hat{\tau}_{2})\Psi_{x}(0-)=(\hat{\tau}_{3}+\mathrm{i}\hat{\tau}_{2})\Psi_{x}(0+)$.
As we know, from this condition, it is obtained that $\psi(0-)=\psi(0+)$
and $\psi_{x}(0-)=\psi_{x}(0+)$. If the relations $\psi_{1}+\psi_{2}=\psi$
(Eq. (5)) and $\psi_{1}-\psi_{2}=(E-V)\psi/\mathrm{m}c^{2}$ (Eq.
(6)) are used (in the latter, we also assumed that $\psi$ is an energy
eigenstate), one can find relations between $\{\Psi(0+),\Psi_{x}(0+)\}$
and $\{\Psi(0-),\Psi_{x}(0-)\}$. We find that the relation given
in Eq. (30) in Ref. \cite{RefU} is none other than the boundary condition
$(\hat{\tau}_{3}+\mathrm{i}\hat{\tau}_{2})\Psi(0-)=(\hat{\tau}_{3}+\mathrm{i}\hat{\tau}_{2})\Psi(0+)$,
with Eqs. (5) and (6) evaluated at $x=0\pm$. Likewise, the relation
given in Eq. (31) of the same reference is none other than $(\hat{\tau}_{3}+\mathrm{i}\hat{\tau}_{2})\Psi_{x}(0-)=(\hat{\tau}_{3}+\mathrm{i}\hat{\tau}_{2})\Psi_{x}(0+)$,
with the spatial derivatives of Eqs. (5) and (6) also evaluated at
$x=0\pm$. Finally, adding the latter two boundary conditions, we
obtain Eq. (32) of Ref. \cite{RefU}. Clearly, if the height of the
step potential is not zero, then $\Psi(0+)$ is different from $\Psi(0-)$,
and $\Psi_{x}(0+)$ is different from $\Psi_{x}(0-)$. Similarly,
in Ref. \cite{RefW}, it was explicitly proven that $\Psi(0+)\neq\Psi(0-)$
and $\Psi_{x}(0+)\neq\Psi_{x}(0-)$ (see Eqs. (19) and (20) in that
reference), but it was also shown that the boundary condition should
be written in the form $(\hat{\tau}_{3}+\mathrm{i}\hat{\tau}_{2})\Psi(0-)=(\hat{\tau}_{3}+\mathrm{i}\hat{\tau}_{2})\Psi(0+)$
and $(\hat{\tau}_{3}+\mathrm{i}\hat{\tau}_{2})\Psi_{x}(0-)=(\hat{\tau}_{3}+\mathrm{i}\hat{\tau}_{2})\Psi_{x}(0+)$.
Incidentally, in the same reference, it was shown that the latter
boundary condition can be obtained by integrating the 1D FV equation
from $x=0-$ to $x=0+$. 

On the other hand, in the problem of the 1D KFG particle inside the
box \textrm{$\Omega=[a,b]$}, and subjected to the potential $V$,
with the Dirichlet boundary condition, $(\hat{\tau}_{3}+\mathrm{i}\hat{\tau}_{2})\Psi(a)=(\hat{\tau}_{3}+\mathrm{i}\hat{\tau}_{2})\Psi(b)=0$,
we know that $\psi$ also satisfies this condition, namely, $\psi(a)=\psi(b)=0$.
The latter boundary condition together with Eqs. (5) and (6) lead
us to the boundary condition $\Psi(a)=\Psi(b)=0$. Indeed, in addition
to $\psi_{1}(a)+\psi_{2}(a)=\psi_{1}(b)+\psi_{2}(b)=0$, $\psi_{1}(a)-\psi_{2}(a)=\psi_{1}(b)-\psi_{2}(b)=0$
(because $\psi_{t}(a,t)=\psi_{t}(b,t)=0$ also holds). Finally, $\Psi$
also satisfies the Dirichlet boundary condition at the edges of the
box (the latter boundary condition was precisely the one used in Ref.
\cite{RefV}). 

In short, let us suppose that the one-component wavefunction $\psi$
can vanish at a point on the real line, for example, at $x=0$ (also
$V(0+)$ and $V(0-)$ must be finite numbers there). The latter is
the Dirichlet boundary condition, namely, $\psi(0-)=\psi(0+)=0\equiv\psi(0)$.
Certainly, this result is obtained from the disappearance of $(\hat{\tau}_{3}+\mathrm{i}\hat{\tau}_{2})\Psi$
at that same point, i.e., from the fact that the Hamiltonian operator
with the latter boundary condition is a pseudo self-adjoint operator;
then, the latter condition implies that the entire two-component wavefunction
$\Psi$ has to disappear at that point (use Eqs. (5) and (6)). In
other words, the 1D FV wave equation is a second-order equation in
the spatial derivative that accepts the vanishing of the entire two-component
wavefunction at a point. On the other hand, let us now suppose that
$\psi_{x}$ can vanish at a point on the real line, for example, at
$x=0$, but $\psi$ is nonzero there (also $V_{x}(0+)$ and $V_{x}(0-)$
must be finite numbers there). The latter is the Neumann boundary
condition, namely, $\psi_{x}(0-)=\psi_{x}(0+)=0\equiv\psi_{x}(0)$.
Indeed, we also have that $(\hat{\tau}_{3}+\mathrm{i}\hat{\tau}_{2})\Psi_{x}$
vanishes at that same point. Then, it can be shown that $(\psi_{1})_{x}$
and $(\psi_{2})_{x}$ do not have to vanish at the point in question,
and therefore, $\Psi_{x}$ is not zero there either (use Eqs. (5)
and (6)). 

\section{Appendix I}

\noindent The 1D KFG wave equation given in Eq. (3) can also be written
as follows:
\[
\left[\,-\hbar^{2}\frac{\partial^{2}}{\partial t^{2}}-\mathrm{i}2\hbar\, V(x)\frac{\partial}{\partial t}+(V(x))^{2}\right]\psi=\left[-\hbar^{2}c^{2}\frac{\partial^{2}}{\partial x^{2}}+(\mathrm{m}c^{2})^{2}\right]\psi,\tag{A1}
\]
and therefore,
\[
\psi_{tt}=c^{2}\psi_{xx}-\left(\frac{\mathrm{m}c^{2}}{\hbar}\right)^{2}\psi+\frac{2V}{\mathrm{i}\hbar}\psi_{t}+\frac{V^{2}}{\hbar^{2}}\psi.\tag{A2}
\]
The scalar product for the two-component column state vectors $\Psi=\left[\,\psi_{1}\;\,\psi_{2}\,\right]^{\mathrm{T}}$
and $\Phi=\left[\,\phi_{1}\;\,\phi_{2}\,\right]^{\mathrm{T}}$, where
$\psi_{1}+\psi_{2}=\psi$ and $\phi_{1}+\phi_{2}=\phi$, is given
by 
\[
\langle\langle\Psi,\Phi\rangle\rangle\equiv\int_{\Omega}\mathrm{d}x\,\Psi^{\dagger}\hat{\tau}_{3}\Phi=\frac{\mathrm{i}\hbar}{2\mathrm{m}c^{2}}\int_{\Omega}\mathrm{d}x\,\left[\psi^{*}\left(\frac{\partial}{\partial t}-\frac{V}{\mathrm{i}\hbar}\right)\phi-\left(\left(\frac{\partial}{\partial t}-\frac{V}{\mathrm{i}\hbar}\right)\psi\right)^{*}\phi\right]
\]
\[
=\frac{\mathrm{i}\hbar}{2\mathrm{m}c^{2}}\int_{\Omega}\mathrm{d}x\,\left(\psi^{*}\phi_{t}-\psi_{t}^{*}\phi-\frac{2V}{\mathrm{i}\hbar}\psi^{*}\phi\right)\equiv\langle\psi,\phi\rangle_{\mathrm{KFG}}.\tag{A3}
\]
The latter quantity is preserved in time; in fact, taking its time
derivative and using the result in Eq. (A2), and a similar relation
for $\phi$ ($\psi$ and $\phi$ are solutions of the 1D KFG wave
equation in its standard form), one obtains the same relation given
in Eq. (14), namely, 
\[
\frac{\mathrm{d}}{\mathrm{d}t}\langle\langle\Psi,\Phi\rangle\rangle=\frac{\mathrm{d}}{\mathrm{d}t}\langle\psi,\phi\rangle_{\mathrm{KFG}}=-\frac{\mathrm{i}\hbar}{2\mathrm{m}}\left.\left[\,\psi_{x}^{*}\,\phi-\psi^{*}\phi_{x}\,\right]\right|_{a}^{b}.\tag{A4}
\]
As follows from the results obtained in Appendix II, if $\psi$ and
$\phi$ both satisfy any boundary condition included in the most general
set of boundary conditions, the boundary term in Eq. (A4) always vanishes.

\section{Appendix II}

\noindent The goal of this section is to show that if the functions
belonging to the domain of $\hat{\mathrm{h}}$ (considered a densely
defined operator) obey any of the boundary conditions included in
Eq. (31), then the functions of the domain of $\hat{\mathrm{h}}_{\mathrm{adj}}$
must obey the same boundary condition. This means that for the general
family of boundary conditions given in Eq. (31), the operator $\hat{\mathrm{h}}=\hat{\mathrm{h}}_{\mathrm{adj}}$
is pseudo self-adjoint. Our results are obtained using simple arguments
that are part of the general theory of linear operators in an indefinite
inner product space (see, for example, Refs. \cite{RefFF,RefGG}).

Let us return to the result given in Eq. (16), namely,
\[
\langle\langle\Xi,\hat{\mathrm{h}}\Phi\rangle\rangle=\langle\langle\hat{\mathrm{h}}_{\mathrm{adj}}\Xi,\Phi\rangle\rangle+f[\Xi,\Phi],\tag{A5}
\]
where $f[\Xi,\Phi]$ is given by (see Eq. (18))
\[
f[\Xi,\Phi]\equiv\frac{\hbar^{2}}{2\mathrm{m}}\,\frac{1}{2}\left.\left[\,\left((\hat{\tau}_{3}+\mathrm{i}\hat{\tau}_{2})\Xi_{x}\right)^{\dagger}(\hat{\tau}_{3}+\mathrm{i}\hat{\tau}_{2})\Phi-\left((\hat{\tau}_{3}+\mathrm{i}\hat{\tau}_{2})\Xi\right)^{\dagger}(\hat{\tau}_{3}+\mathrm{i}\hat{\tau}_{2})\Phi_{x}\,\right]\right|_{a}^{b}.\tag{A6}
\]
Here, $\hat{\mathrm{h}}$ can act on column vectors $\Phi=\left[\,\phi_{1}\;\,\phi_{2}\,\right]^{\mathrm{T}}\in\mathcal{D}(\hat{\mathrm{h}})$,
where $\mathcal{D}(\hat{\mathrm{h}})$ is the domain of $\hat{\mathrm{h}}$,
a set of column vectors on which we allow the differential operator
$\hat{\mathrm{h}}$ to act ($\mathcal{D}(\hat{\mathrm{h}})$ is a
linear subset of the indefinite inner product space), which fundamentally
includes boundary conditions, and $\hat{\mathrm{h}}_{\mathrm{adj}}$
can act on column vectors $\Xi=\left[\,\xi_{1}\;\,\xi_{2}\,\right]^{\mathrm{T}}\in\mathcal{D}(\hat{\mathrm{h}}_{\mathrm{adj}})$
(in general, $\mathcal{D}(\hat{\mathrm{h}}_{\mathrm{adj}})$ may not
coincide with $\mathcal{D}(\hat{\mathrm{h}})$). By virtue of the
result given in Eq. (5), the respective solutions of Eq. (3) are the
following:
\[
\phi_{1}+\phi_{2}=\phi\quad\mathsf{and}\quad\xi_{1}+\xi_{2}=\xi.\tag{A7}
\]
The boundary term in Eq. (A6) can be written in terms of $\phi$ and
$\xi$, namely,
\[
f[\Xi,\Phi]=\frac{\hbar^{2}}{2\mathrm{m}}\left.\left[\,\xi_{x}^{*}\,\phi\,-\xi^{*}\phi_{x}\,\right]\right|_{a}^{b}.\tag{A8}
\]

First, let us suppose that every column vector $\Phi\in\mathcal{D}(\hat{\mathrm{h}})$
satisfies the boundary conditions $(\hat{\tau}_{3}+\mathrm{i}\hat{\tau}_{2})\Phi(a)=(\hat{\tau}_{3}+\mathrm{i}\hat{\tau}_{2})\Phi(b)=0$
and $(\hat{\tau}_{3}+\mathrm{i}\hat{\tau}_{2})\Phi_{x}(a)=(\hat{\tau}_{3}+\mathrm{i}\hat{\tau}_{2})\Phi_{x}(b)=0$,
or, equivalently, $\phi(a)=\phi(b)=0$ and $\phi_{x}(a)=\phi_{x}(b)=0$
(remember the first relation in Eq. (A7)). In this case, the boundary
term in Eq. (A5) vanishes, and we have the result
\[
\langle\langle\Xi,\hat{\mathrm{h}}\Phi\rangle\rangle=\langle\langle\hat{\mathrm{h}}_{\mathrm{adj}}\Xi,\Phi\rangle\rangle.\tag{A9}
\]
The latter relation is precisely the one that defines the generalized
adjoint differential operator. It is clear that its verification did
not require the imposition of any boundary condition on the vectors
$\Xi\in\mathcal{D}(\hat{\mathrm{h}}_{\mathrm{adj}})$. Thus, until
now, we have that $\mathcal{D}(\hat{\mathrm{h}})\neq\mathcal{D}(\hat{\mathrm{h}}_{\mathrm{adj}})$
(in fact, we have that $\mathcal{D}(\hat{\mathrm{h}})\subset\mathcal{D}(\hat{\mathrm{h}}_{\mathrm{adj}})$,
i.e., $\mathcal{D}(\hat{\mathrm{h}})$ is a restriction of $\mathcal{D}(\hat{\mathrm{h}}_{\mathrm{adj}})$). 

If the operator $\hat{\mathrm{h}}$ is to be a pseudo self-adjoint
differential operator, the relation given in Eq. (21), namely, $\hat{\mathrm{h}}=\hat{\mathrm{h}}_{\mathrm{adj}}$,
must be verified, and therefore, $\mathcal{D}(\hat{\mathrm{h}})=\mathcal{D}(\hat{\mathrm{h}}_{\mathrm{adj}})$.
To achieve this, we must allow every vector $\Phi\in\mathcal{D}(\hat{\mathrm{h}})$
to satisfy more general boundary conditions, that is, we must relax
the domain of $\hat{\mathrm{h}}$. Let us suppose that we have a set
of boundary conditions to be imposed on a vector $\Phi\in\mathcal{D}(\hat{\mathrm{h}})$;
if the cancellation of the boundary term $f[\Xi,\Phi]$ by these boundary
conditions only depends on imposing the same boundary conditions on
the vector $\Xi\in\mathcal{D}(\hat{\mathrm{h}}_{\mathrm{adj}})$,
then $\hat{\mathrm{h}}$ will be a pseudo self-adjoint differential
operator.

First, from Eq. (A8), we write the boundary term in Eq. (A5) as follows:
\[
\lambda\frac{2\mathrm{m}}{\hbar^{2}}f[\Xi,\Phi]=\left.\left[\,\phi\,\lambda\xi_{x}^{*}-\xi^{*}\lambda\phi_{x}\,\right]\right|_{a}^{b}
\]
\[
=\left[\,\phi(b)\,\lambda\xi_{x}^{*}(b)-\xi^{*}(b)\,\lambda\phi_{x}(b)\,\right]-\left[\,\phi(a)\,\lambda\xi_{x}^{*}(a)-\xi^{*}(a)\,\lambda\phi_{x}(a)\,\right]=0.\tag{A10}
\]

\noindent It is fairly convenient to rewrite the latter two terms
using the following identity:
\[
z_{1}z_{2}^{*}-z_{3}^{*}z_{4}=\frac{\mathrm{i}}{2}\left[\,(z_{1}+\mathrm{i}z_{4})(z_{3}+\mathrm{i}z_{2})^{*}-(z_{1}-\mathrm{i}z_{4})(z_{3}-\mathrm{i}z_{2})^{*}\,\right],\tag{A11}
\]
where $z_{1}$, $z_{2}$, $z_{3}$ and $z_{4}$ are complex numbers.
The latter relation is the generalization of that given in Eq. (27).
In fact, making the replacements $z_{3}\rightarrow z_{1}$ and $z_{4}\rightarrow z_{2}$
in Eq. (A11), the relation given in Eq. (27) is obtained. Then, the
following result is derived:
\[
\lambda\frac{2\mathrm{m}}{\hbar^{2}}f[\Xi,\Phi]=\frac{\mathrm{i}}{2}\left[\left(\phi(b)+\mathrm{i}\lambda\phi_{x}(b)\right)\left(\xi(b)+\mathrm{i}\lambda\xi_{x}(b)\right)^{*}-\left(\phi(b)-\mathrm{i}\lambda\phi_{x}(b)\right)\left(\xi(b)-\mathrm{i}\lambda\xi_{x}(b)\right)^{*}\right]
\]
\[
-\frac{\mathrm{i}}{2}\left[\left(\phi(a)+\mathrm{i}\lambda\phi_{x}(a)\right)\left(\xi(a)+\mathrm{i}\lambda\xi_{x}(a)\right)^{*}-\left(\phi(a)-\mathrm{i}\lambda\phi_{x}(a)\right)\left(\xi(a)-\mathrm{i}\lambda\xi_{x}(a)\right)^{*}\right]
\]
\[
=\frac{\mathrm{i}}{2}\left[\left(\phi(b)+\mathrm{i}\lambda\phi_{x}(b)\right)\left(\xi(b)+\mathrm{i}\lambda\xi_{x}(b)\right)^{*}+\left(\phi(a)-\mathrm{i}\lambda\phi_{x}(a)\right)\left(\xi(a)-\mathrm{i}\lambda\xi_{x}(a)\right)^{*}\right]
\]
\[
-\frac{\mathrm{i}}{2}\left[\left(\phi(b)-\mathrm{i}\lambda\phi_{x}(b)\right)\left(\xi(b)-\mathrm{i}\lambda\xi_{x}(b)\right)^{*}+\left(\phi(a)+\mathrm{i}\lambda\phi_{x}(a)\right)\left(\xi(a)+\mathrm{i}\lambda\xi_{x}(a)\right)^{*}\right]=0,
\]
this means that
\[
\lambda\frac{2\mathrm{m}}{\hbar^{2}}f[\Xi,\Phi]=\frac{\mathrm{i}}{2}\left[\begin{array}{c}
\xi(b)+\mathrm{i}\lambda\xi_{x}(b)\\
\xi(a)-\mathrm{i}\lambda\xi_{x}(a)
\end{array}\right]^{\dagger}\left[\begin{array}{c}
\phi(b)+\mathrm{i}\lambda\phi_{x}(b)\\
\phi(a)-\mathrm{i}\lambda\phi_{x}(a)
\end{array}\right]
\]
\[
-\frac{\mathrm{i}}{2}\left[\begin{array}{c}
\xi(b)-\mathrm{i}\lambda\xi_{x}(b)\\
\xi(a)+\mathrm{i}\lambda\xi_{x}(a)
\end{array}\right]^{\dagger}\left[\begin{array}{c}
\phi(b)-\mathrm{i}\lambda\phi_{x}(b)\\
\phi(a)+\mathrm{i}\lambda\phi_{x}(a)
\end{array}\right]=0.\tag{A12}
\]

Let us now consider a more general set of boundary conditions to be
imposed on a vector $\Phi\in\mathcal{D}(\hat{\mathrm{h}})$ (i.e.,
more general than the boundary conditions that we presented after
Eq. (A8)), namely, 
\[
\left[\begin{array}{c}
\phi(b)+\mathrm{i}\lambda\phi_{x}(b)\\
\phi(a)-\mathrm{i}\lambda\phi_{x}(a)
\end{array}\right]=\hat{\mathrm{N}}\left[\begin{array}{c}
\phi(b)-\mathrm{i}\lambda\phi_{x}(b)\\
\phi(a)+\mathrm{i}\lambda\phi_{x}(a)
\end{array}\right],\tag{A13}
\]
where $\hat{\mathrm{N}}$ in an arbitrary complex matrix. By substituting
the latter relation in Eq. (A12), we obtain the following result:
\[
\lambda\frac{2\mathrm{m}}{\hbar^{2}}f[\Xi,\Phi]
\]
\[
=\frac{\mathrm{i}}{2}\left\{ \left(\left[\begin{array}{c}
\xi(b)+\mathrm{i}\lambda\xi_{x}(b)\\
\xi(a)-\mathrm{i}\lambda\xi_{x}(a)
\end{array}\right]^{\dagger}\hat{\mathrm{N}}-\left[\begin{array}{c}
\xi(b)-\mathrm{i}\lambda\xi_{x}(b)\\
\xi(a)+\mathrm{i}\lambda\xi_{x}(a)
\end{array}\right]^{\dagger}\right)\left[\begin{array}{c}
\phi(b)-\mathrm{i}\lambda\phi_{x}(b)\\
\phi(a)+\mathrm{i}\lambda\phi_{x}(a)
\end{array}\right]\right\} =0,
\]
and therefore,
\[
\left[\begin{array}{c}
\xi(b)+\mathrm{i}\lambda\xi_{x}(b)\\
\xi(a)-\mathrm{i}\lambda\xi_{x}(a)
\end{array}\right]^{\dagger}\hat{\mathrm{N}}=\left[\begin{array}{c}
\xi(b)-\mathrm{i}\lambda\xi_{x}(b)\\
\xi(a)+\mathrm{i}\lambda\xi_{x}(a)
\end{array}\right]^{\dagger}\tag{A14}
\]
(This result is because, at this point, we cannot impose any boundary
conditions that would completely annul the column vectors in Eq. (A13),
for example). Every vector $\Xi\in\mathcal{D}(\hat{\mathrm{h}}_{\mathrm{adj}})$
should satisfy the same boundary conditions that $\Phi\in\mathcal{D}(\hat{\mathrm{h}})$
satisfies, i.e., the boundary conditions in Eq. (A13), namely, 
\[
\left[\begin{array}{c}
\xi(b)+\mathrm{i}\lambda\xi_{x}(b)\\
\xi(a)-\mathrm{i}\lambda\xi_{x}(a)
\end{array}\right]=\hat{\mathrm{N}}\left[\begin{array}{c}
\xi(b)-\mathrm{i}\lambda\xi_{x}(b)\\
\xi(a)+\mathrm{i}\lambda\xi_{x}(a)
\end{array}\right].\tag{A15}
\]
Taking the Hermitian conjugate of the matrix relation in Eq. (A14)
and substituting this result into Eq. (A15), we obtain
\[
\left[\begin{array}{c}
\xi(b)+\mathrm{i}\lambda\xi_{x}(b)\\
\xi(a)-\mathrm{i}\lambda\xi_{x}(a)
\end{array}\right]=\hat{\mathrm{N}}\hat{\mathrm{N}}^{\dagger}\left[\begin{array}{c}
\xi(b)+\mathrm{i}\lambda\xi_{x}(b)\\
\xi(a)-\mathrm{i}\lambda\xi_{x}(a)
\end{array}\right];
\]
therefore, $\hat{\mathrm{N}}$ is a unitary matrix. Thus, the most
general family of pseudo self-adjoint, or generalized self-adjoint
boundary conditions, for the 1D KFG particle in a box can be written
in the form given by Eq. (31), namely, 
\[
\left[\begin{array}{c}
\xi(b)-\mathrm{i}\lambda\xi_{x}(b)\\
\xi(a)+\mathrm{i}\lambda\xi_{x}(a)
\end{array}\right]=\hat{\mathrm{U}}\left[\begin{array}{c}
\xi(b)+\mathrm{i}\lambda\xi_{x}(b)\\
\xi(a)-\mathrm{i}\lambda\xi_{x}(a)
\end{array}\right],\tag{A16}
\]
where $\hat{\mathrm{U}}=\hat{\mathrm{N}}^{-1}$. The fact that the
boundary condition for $\Phi\in\mathcal{D}(\hat{\mathrm{h}})$ (for
example, given in terms of $\phi$) is the same boundary condition
for $\Xi\in\mathcal{D}(\hat{\mathrm{h}}_{\mathrm{adj}})$ (given in
terms of $\xi$) ensures that $\mathcal{D}(\hat{\mathrm{h}})=\mathcal{D}(\hat{\mathrm{h}}_{\mathrm{adj}})$;
therefore, $\hat{\mathrm{h}}$, which was already a pseudo-Hermitian
operator, is also a pseudo self-adjoint operator. Additionally, the
boundary term given in Eq. (14), or in Eq. (A4), vanishes, and therefore,
the pseudo inner product is conserved. 

\section{Concluding remarks}

\noindent The KFG Hamiltonian operator, or the Hamiltonian that is
present in the first order in time 1D KFG wave equation, i.e., the
1D FV wave equation, is formally pseudo-Hermitian. This is a well-known
fact, and its verification does not require knowledge of the domain
of the Hamiltonian or its adjoint. We have shown that this operator
is also a pseudo-Hermitian operator, but in addition, it is a pseudo
self-adjoint operator when it describes a 1D KFG particle in a finite
interval. Consequently, we constructed the most general set of boundary
conditions for this operator, which is characterized by four real
parameters and is consistent with the last two properties. All these
results can be extended to the problem of a 1D KFG particle moving
on a real line with a penetrable or an impenetrable obstacle at one
point, i.e., with a point interaction (or a hole) there. For instance,
assuming the point is $x=0$, it suffices to make the replacements
$x=a\to0+$ and $x=b\to0-$ in the general set of boundary conditions
for the particle in the interval $[a,b]$. 

As we have shown, the general set of boundary conditions can be written
in terms of the one-component wavefunction for the second order in
time 1D KFG wave equation, that is, $\psi$, and its derivative $\psi_{x}$,
both evaluated at the ends of the box. Certainly, we showed that the
general set can also be written in terms of the two-component column
vectors for the 1D FV wave equation, that is, $(\hat{\tau}_{3}+\mathrm{i}\hat{\tau}_{2})\Psi$
and $(\hat{\tau}_{3}+\mathrm{i}\hat{\tau}_{2})\Psi_{x}$, evaluated
at the ends of the box. We only used algebraic arguments and simple
concepts that are within the general theory of linear operators on
a space with indefinite inner product to build these sets of boundary
conditions.

From the results presented in Section III, we also found that $\Psi$
and $\Psi_{x}$ do not necessarily satisfy the same boundary condition
that $(\hat{\tau}_{3}+\mathrm{i}\hat{\tau}_{2})\Psi$ and $(\hat{\tau}_{3}+\mathrm{i}\hat{\tau}_{2})\Psi_{x}$
satisfy. In any case, given a particular boundary condition that $\psi$
and $\psi_{x}$ satisfy at the ends of the box and using the relations
that arise between the components of the column vector $\Psi$, that
is, $\psi_{1}$ and $\psi_{2}$, and quantities $\psi$, $\psi_{t}$,
and the potential $V$ (see Eqs. (5) and (6)), the respective boundary
condition on $\Psi$ and $\Psi_{x}$ can be obtained. 

We think that our article will be of interest to those interested
in the fundamental and technical aspects of relativistic wave equations.
Furthermore, to the best of our knowledge, the main results of our
article, i.e., those related to general pseudo self-adjoint sets of
boundary conditions in the 1D KFG theory, do not appear to have been
considered before. 

\section*{Acknowledgments}

\noindent The author wishes to thank the referees for their comments and suggestions.

\section*{Conflicts of interest}

\noindent The author declares no conflicts of interest.

\end{document}